\documentclass[12pt]{article}
\usepackage{amssymb}
\usepackage{amsmath}
\usepackage{graphicx}

\begin{document}

\vspace*{1cm}

\begin{center} 
\setlength{\baselineskip}{24pt}
{\LARGE A new method for the determination \\
of the real part  of the hadron elastic scattering\\
amplitude at small angles and high energies}
\end{center}

\begin{center}
\vspace{2cm}
{\large P. Gauron, B. Nicolescu}

\vspace{0.5cm} 
 Theory Group,  Laboratoire de Physique Nucl\'eaire  et des Hautes  \'Energies 
(LPNHE)\footnote{Unit\'e  de Recherche des Universit\'es 
  Paris 6 et Paris 7, Associ\'ee ou CNRS},
 CNRS and Universit\'e Pierre et Marie Curie, Paris\\ 
{\small e-mail: \texttt{gauron@in2p3.fr} }\\
     {\small e-mail: \texttt{nicolesc@lpnhep.in2p3.fr }}

\vspace{1cm}
{\large O.V. \ Selyugin}  

\vspace{0.5cm} 
 BLTP, JINR and Universit\'e  de Li\`ege   \\
    {\small email: \texttt{selugin@thsun1.jinr.ru}} 

\vspace{3cm}
\textbf{Abstract}
\end{center}
A new method for the determination of the real part of  the elastic scattering amplitude  is examined for high energy proton-proton at small momentum transfer. 
This method allows us to decrease the number of  model assumptions,  to obtain the real part  in a narrow region  of momentum transfer and to test  different models.
The real part is computed at a given point $t_{min}$ near $t=0$ from the known Coulomb amplitude.
Hence one obtains an important constraint on the real part of the forward scattering amplitude and therefore on the $\rho$-parameter (measuring the ratio of the real to imaginary part of the scattering amplitude at $t=0$), which can be tested at LHC.

\section{Introduction - The standard method for extracting real parts}

The standard procedure to extract the magnitude of the real part of the hadron elastic scattering includes a fit to the experimental data in the interference region, by minimizing the $\chi^2$ function:
\begin{equation}
\label{fit}
 \chi^2\!=\!\sum_{i=1}^{k} \frac{(d\sigma^{\mbox{\scriptsize exp}}/ dt(t\!=\!t_i)
              \!  -\!d\sigma/ dt(t\!=\!t_i) )^2}{\Delta^{2}_{\mbox{\scriptsize exp, i}}} ,
\end{equation}
where the experimental differential cross section $d\sigma^{\mbox{\scriptsize exp}}/ dt(t=t_i)$ at the point $t_i$  and the statistical error $\Delta_{\mbox{\scriptsize exp, i}}$ are extracted from the measured $dN/dt$ using, for example, the value of the luminosity.

The theoretical representation of the differential cross-sections is
\begin{equation}
\label{dsig}
{d\sigma\over dt}= 2 \pi
[\vert \Phi_1\vert^2+\vert \Phi_2\vert^2
    + \vert \Phi_3\vert^2+\vert \Phi_4\vert^2+4\vert
 \Phi_5\vert^2]\ ,
\end{equation}
where  $\Phi_1$ and $\Phi_3$ are the spin non-flip amplitudes.
The total helicity amplitudes can be written as a sum of nuclear $\Phi_i^h(s,t)$ and electromagnetic  $\Phi_i^e(s,t)$ amplitudes:
\begin{equation}
\label{phii}
\Phi_i(s,t)=\Phi_i^h(s,t)+e^{i\alpha\varphi}\Phi_i^e(t)\ ,
\end{equation}
where $\Phi_i^e(t)$ are the leading terms at high energies for the one-photon amplitude as defined, for example, in \cite{leader} and $\alpha$ is the fine-structure constant. 
The common phase $\varphi$ is 
\begin{equation}
\label{varphi}
\varphi=\mp[\gamma+\log\big(B(s,t)\vert t\vert/2\big)+\nu_1+\nu_2],
\end{equation}
where the upper (lower) sign is related to the $pp$ ($p\bar p$) scattering, $B(s,t)$ is the slope of the differential cross section, $\gamma$ is the Euler constant ($\gamma= 0.577...$) and $\nu_1$ and $\nu_2$ are small correction terms  defining the behavior of the Coulomb-hadron phase at small momentum transfers (see \cite{Cahn} and, more recently, \cite{selprd}).
At very small $t$ and fixed $s$, the electromagnetic amplitudes are such that
$
\Phi_1^e(t) = \Phi_3^e(t)\sim\alpha/t \ ,
\Phi_2^e(t) = -\Phi_4^e(t)\sim \alpha\cdot \hbox{const.}\ ,
\Phi_5^e(t)  \sim  -\alpha/\sqrt{\vert t\vert}\ .
$
We assume, as usual, that at high energies and small angles the one-flip and double-flip hadron amplitudes are small with respect to the spin-nonflip ones and that the hadron contributions to $\Phi_1$ and $\Phi_3$ are the same, as are the electromagnetic ones. Therefore
\begin{equation}
\label{sup}
F (s,t) = F_{N}+F_{C} \ \exp( i \alpha \varphi).
\end{equation}
In the $O(\alpha)$ approximation, one has:
\begin{equation}
   d\sigma/dt     =  \pi |e^{i \alpha \varphi} F_C + F_N|^2  
   =\pi \ [ (  F_C + Re F_N)^2
               +(\alpha \varphi F_C + Im F_N)^2] \ .
\label{ds0}
\end{equation}
In the standard fitting procedure, one neglects the $\alpha^2$ term in Eq.~(\ref{ds0}) and this equation takes the form:
\begin{equation}
d\sigma/dt = \pi [ (F_{C} (t))^2\!+\! (\rho(s,t)^2 + 1) (Im F_{N}(s,t))^{2})    
+ 2 (\rho(s,t)+ \alpha \varphi(t)) F_{C}(t) Im F_{N}(s,t)],
\label{ds2}
\end{equation}
where $F_{C}(t) = \mp\ 2 \alpha G^{2}(t)/|t|$ is the Coulomb amplitude (the upper sign is for $pp$, the lower sign is for $p\bar p$) and $G^2(t)$ is  the  proton electromagnetic form factor squared;
$ReF_N(s,t)$ and $ImF_N(s,t)$ are the real and imaginary parts of the hadron amplitude;
$\rho(s,t) = Re F_N(s,t) / Im F_N(s,t)$.
The formula (\ref{ds2}) is used for the fit of experimental  data in getting hadron amplitudes and the Coulomb-hadron phase in order to obtain the value of $\rho(s,t)$.

\section{Computation of the real part of the spin-non-flip amplitude of the $pp$ scattering from the Coulomb amplitude at a given point $t_{min}$ near $t=0$} 

Numerous  discussions  of the $\rho$-parameter (the value of $\rho(s,t)$ at $t=0$) measured by the UA4 \cite{ua4} and UA4/2 \cite{ua42} Collaborations in $p\bar p$ scattering at $\sqrt{s}=541$ GeV have revealed the ambiguity in the definition of this semi-theoretical parameter.
As a result, it has been shown that one has some trouble in extracting from experiment the total cross sections and the value of the forward ($t=0$) real part of the scattering amplitudes \cite{selpl}-\cite{gns97}.
In fact, the problem is that we have at our disposal only one observable $d\sigma/dt$ for two unknowns, the real  and imaginary parts of $  F_{N}(s,t)$.
So, we need either some additional experimental information which would allow us to determine independently the real and imaginary parts of the spin non-flip hadron elastic scattering amplitude or some new ways to determine the magnitude of the phase of the hadron scattering amplitude with a minimum number of theoretical assumptions.

Let us note two points concerning the familiar exponential forms of $ReF_N(s,t)$ and $ImF_N(s,t)$ used by experimentalists. 
First, for simplicity reasons, one makes the assumption that the slope of imaginary part  of the scattering amplitude is equal to the slope of its real part in  the examined range of momentum transfer, and, for the best fit, one should take the interval of momentum transfer sufficiently large.
Second, the magnitude of $\rho(s,t)$ thus obtained  corresponds to the whole interval of momentum transfer.

In this article, we briefly describe new and more general procedures simplifying the determination of elastic scattering amplitude parameters.

From equation  (\ref{ds0}), one can obtain an equation for  $Re F_{N}(s,t)$ for every experimental point  $i$: 
\begin{equation}
Re F_{N}(s,t_i)= - F_{C}(t_i)   
 \pm [(1/ \pi) d\sigma^{exp}/ dt(t_i)
 - (\alpha \varphi F_{C}(t_i)+Im F_{N}(s,t_i))^2]^{1/2}\ .
 \label{rsq}
\end{equation}

The experimentalists define the imaginary part of the scattering amplitude {\it via} the usual exponential approximation in the small $t$-region
\begin{equation}
\label{im}
Im F_{N}(s,t) = \sigma_{tot}/(0.389 \cdot 4 \pi) \exp(B t/2),
\end{equation}
 where 0.389 is the usual converting dimensional factor for expressing $\sigma_{\mbox{tot}}$ in mb.
 
Equation (\ref{rsq}) shows  the possibility to calculate the real part at every separate point $t_i$ if the imaginary part of the  scattering amplitude is fixed and to check the exponential form of the obtained  real part of the scattering amplitude (see \cite{selyf}).

Let us define the sum of the real parts of the hadron and Coulomb amplitudes as $\sqrt{\Delta_{R}}$, so we can write:
\begin{equation}
\label{Del}
 \Delta^{th}_{R}(s,t_i) =  [ Re F_{N}(s,t_i)+ F_{C}(t_i)]^2\geq 0\ .
\end{equation}
Using  the differential cross sections experimental data we obtain:
\begin{equation}
\label{Del2}
\Delta^{exp}_{R}(s,t_i)  =   (1/\pi) \  d\sigma^{exp}/ dt(t_{i}) 
- ( \alpha \varphi F_{C}(t_i)+Im F_{N}(s,t_i))^2\ .
\end{equation}
For  $pp$ scattering at high energies, eqs.~(\ref{Del}) and (\ref{Del2})  induce  a remarkable property.
Let us note that the real part of the Coulomb $pp$ scattering amplitude is negative and exceeds the size of $F_N^{pp} (s,t)$ at $t \rightarrow 0$, but has a large slope.
As the real part of the hadron amplitude is known as being positive at relatively high (ISR) energies,
it is obvious that  $\Delta_R^{th}$ must go through zero at some value $t=t_{min}^{pp}$, i. e.
\begin{equation}
\label{ReFN}
ReF_N^{pp}(t_{min}^{pp})=-F_C^{pp}(t_{min}^{pp})
\end{equation}
and
\begin{equation}
\label{DthR}
\Delta_R^{th}(s,t_{min}^{pp})=0\ .
\end{equation}
Therefore $\Delta_R^{exp}$ must have a minimum at the same value $t=t_{min}^{pp}$.

The interpretation of Eq.~(\ref{ReFN}) is obvious: \textit{at fixed} $s$, \textit{the real part of the $pp$ amplitude is computed from the Coulomb amplitude at} $t=t_{min}^{pp}$.

The magnitude of $\Delta_R^{exp}(s,t)$ as compared with $\Delta_R^{th}(s,t)$ gives as a measure of the accuracy of the experiment and of the theoretical model : $\Delta_R^{exp}(s,t)$ has to be very close to 
$\Delta_R^{th}(s,t)$.
Consequently, $\Delta_R^{exp}(s,t_{min}^{pp})$ should be almost zero.
The value of $t_{min}^{pp}$, defined in Eq. (\ref{DthR}), is determined, in fact, by the minimum of 
$\Delta_R^{exp}$.
If the position $t_{min}^{pp}$ of the minimum of $\Delta_R^{exp}$ is different from the position of the zero of $\Delta_R^{th}$, then the model is questionable as concerns $ReF_N^{pp}$.
This gives a powerful test for any model.

Namely, in the case of the exponential forms, we have
\begin{eqnarray}
\label{rhost}
\rho^{pp}(s,t)& = & \frac{ReF_N^{pp}(s,t)}{ImF_N^{pp}(s,t)}=\rho^{pp}(s,0)\nonumber \\
& = & \mbox{ const }=\rho^{pp}(s,t_{min})
\end{eqnarray} 
However our method gives the possibility to extract $\rho^{pp}(s,t_{min}^{pp})$ \textit{without} assuming the exponential form for $ReF_N^{pp}(s,t)$, from eqs.~(\ref{im}) and (\ref{ReFN}). 
If this numerical value of $\rho^{pp}(s,t_{min}^{pp})$ is significantly different from the value $\rho^{pp}(s,0)$ extracted by a given experiment, this means that the exponential form of $ReF_N^{pp}(s,t)$ is doubtful.

Our method gives the possibility to extract the real part $ReF_N^{pp}(s,t)$ at $t=t_{min}^{pp}$ without assuming neither an exponential form nor any other specific form for the real part.
Moreover, we know  (e.g. from the Regge model) that the forward hadron scattering amplitude is predominantly imaginary.
Therefore a model which describes well the experimental $dN/dt$ data necessarily has a good $ImF_N(s,t)$ for high $s$ and small $t$. 
Our method precisely uses a given model for $ImF_N^{pp}$ which is supposed to describe well the experimental data.

In other words, our method is quasi model-independent : different models for $ImF_N(s,t)$ lead to a quite restricted range of values of $t_{min}$.

Let us underline, in order to avoid any misunderstandings, that our method is by no means aimed to extract $ReF_N(s,t)$ as a function of $t$ for a given $s$.
Because of the dynamical dominance of the imaginary part of the hadron amplitude (described mainly by the Pomeron) over the real part, the expression of $\Delta_R^{exp}(s,t_i)$ involves delicate cancellations between the two terms in Eq. (\ref{Del2}) and therefore one deals effectively with small quantities affected by large errors.
Nothing precise about $ReF_N(s,t)$ as a function of $t$ could be said before doing detailed and constrained fits of the data.
Such fits are beyond the aim of the present paper.
Our aim is to impose as a constraint for all existing models the zero in $\Delta_R(s,t_i)$ which leads to a rather precise value of $ReF_N(s,t)$ at \textit{a special point} $t=t_{min}$, value computed from the Coulomb amplitude at the same special point.
Even such a restricted calculation requires high-precision data and a large number of experimental points.
The problem here is that we extract a small quantity - the real part of the hadron elastic amplitude - affected by large errors. In order to minimize these errors we need a very high-precision experiment.
The only $pp$ data we did find in literature, satisfying our criterium, are those at $\sqrt{s}=52.8$ GeV \cite{528}.
We therefore pedagogically illustrate how our method works by taking the case of these data.

In Fig. 1 we plot $\Delta_R^{exp}(s,t_i)$ as given by eqs. (\ref{Del2}) and (\ref{im}), with $\sigma_T^{pp}=42.38$ mb and $B^{pp}=12.87 \mbox{ (GeV)}^{-2}$ \cite{528}. The error bars of the $\Delta_R^{exp}$ points are calculated from the errors bars of $d\sigma^{exp}/dt$ points.


We also plot on the same figure $\Delta_R^{th}(s,t_i)$ as given by Eq. (\ref{Del}), where 
\begin{equation}
\label{new15}
ReF_N^{pp}(s,t)=(\rho^{pp}\cdot\sigma_{tot}^{pp})/(0.389\cdot 4\pi)\exp(B^{pp}t/2),
\end{equation}
with $\rho^{pp}=0.077$.

We see from Fig. 1 that there is a clear disagreement between $\Delta_R^{th}(s,t_i)$ and $\Delta_R^{exp}(s,t_i)$ in the region
\begin{equation}
\label{new16}
0.03<-t<0.06 \mbox{ GeV}^2\ .
\end{equation}
Namely, $\Delta_R^{th}(s,t_i)$ goes through zero at $-t\simeq 0.024  \mbox{ GeV}^2$ while $\Delta_R^{exp}(s,t_i)$ goes through a minimum at a very different value of $t$. 
Moreover, the values of the two quantities are very different in the region (\ref{new16}).

In fact the entire shape of $\Delta_R^{th}$ in the region (\ref{new16}) is not consistent with the shape of 
$\Delta_R^{exp}$. As it can be seen from Fig. 1, $\Delta_R^{th}$ rises very slowly in the region (\ref{new16}), while $\Delta_R^{exp}$ shows a rapid rise in this region\footnote{The negativity of several points of $\Delta_R^{exp}$ (see Fig. 1) is not important for our discussion. A very small correction of the normalization factor, taking into account systematical errors, and/or of the model used for $ImF_N$, eliminate this negativity.}.

In order to see if this discrepancy is significant we define the corresponding $\chi^2$ value:
\begin{equation}
\label{new17}
\chi^2\vert_{\Delta_R}=\sum_{i=1}^k\frac{\left(\Delta_R^{exp}(s,t_i)-\Delta_R^{th}(s,t_i)\right)^2}{\delta^2\left(\Delta_R^{exp}(s,t_i)\right)},
\end{equation}
where $\delta$ denotes the statistical error of $\Delta_R^{exp}$. 
The overall $\chi^2/pt$ value is 2.4 for a total of 34 points. However the major contribution to $\chi^2$ comes from the region (\ref{new16}), i.e. from only 10 points. Namely the value of $\chi^2/pt$ for the first 24 points is 1.2 while the value of the $\chi^2/pt$ for the last 10 points is 5.2.
The effect shown in Fig. 1 is clearly statistically significant and can not be due to a statistical fluctuation.

We can easily retrace the origin of the effect to $d\sigma/dt$ itself, because of the obvious equality
\begin{equation}
\label{new18}
\chi^2\vert_{\Delta_R}=\chi^2\vert_{d\sigma/dt}\ ,
\end{equation}
which signifies that $d\sigma/dt$ is not well fitted in the region [\ref{new16}].
In order to illustrate the effect in $d\sigma/dt$, we plot in Fig. 2 the quantity
\begin{equation}
\label{new19}
r\equiv \frac{d\sigma^{exp}/dt}{d\sigma^{th}/dt}-1,
\end{equation}
where we take as a theoretical model "th" the exponential model defined by eqs. (\ref{im}) and (\ref{new15}).
The quantity $r$ is clearly different from 0 in the region (\ref{new16}).


In order to evaluate the $t_{min}^{pp}$ value we performed a polynomial fit of $\Delta_R^{exp}(s,t_i)$ with the form
\begin{equation}
\label{new20}
\Big[
a_1\cdot\vert t\vert^{-3/2}+a_2\cdot\vert t\vert^{-2}+a_3\cdot\vert t\vert^{-1}
 +a_4+a_5\cdot\vert t\vert\Big](\vert t\vert-a_6)\ .
\end{equation}
We get a $\chi^2/pt$ value of 0.73, for the following set of parameters (all parameters are expressed in GeV$^{-4}$ ; a scale factor $t_0=1\mbox{ GeV}^2$ is implicitely supposed everywhere in Eq. (\ref{new20})) :
\begin{equation}
\label{new21}
\begin{array}{rclrcl}
   a_1 & = & -0.08649, & a_4 &= & -105.8709,\\
    a_2 & = & -0.00311,   &  a_5 & = & 3154.11,  \\
    a_3 & = & 1.68189,  & a_6 & = & 0.04508.    
\end{array}
\end{equation}
The result of this polynomial fit is shown in Fig. 3.
The corresponding value of $t_{min}^{pp}$ is
\begin{equation}
\label{new22}
t_{min}^{pp}=-0.0325 \pm 0.0025\mbox{ GeV}^2,
\end{equation}
significantly different from the value $t=-0.024\mbox{ GeV}^2$ where $\Delta_R^{th}(s,t_i)$ goes through zero.


We can therefore evaluate, from Eq. (\ref{ReFN}),
\begin{equation}
\label{new23}
ReF_N^{pp}(\sqrt{s}=52.8 \mbox{ GeV},\ t=t_{min}^{pp})
=0.375\pm 0.037 \mbox{ (GeV)}^{-2}
\end{equation}
and, from Eq. (\ref{im}),
\begin{equation}
\label{new24}
ImF_N^{pp} (\sqrt{s}=52.8 \mbox{ GeV},\ t=t_{min}^{pp})
=7.027  \mbox{ (GeV)}^{-2}\ .
\end{equation}
Therefore
\begin{equation}
\label{new25}
\rho^{pp} (\sqrt{s}=52.8 \mbox{ GeV},\ t=t_{min}^{pp})
=0.053\pm 0.005\ ,
\end{equation}
a value which is somewhat different ($\sim 2$ standard deviations) from the value given in Ref. \cite{528}:
\begin{equation}
\label{new26}
\rho^{pp} (\sqrt{s}=52.8 \mbox{ GeV},\ t=t_{min}^{pp})
=0.077\pm 0.009\ .
\end{equation} 
The difference between the $\rho$-values, expressed by (\ref{new25}) and (\ref{new26}), is not highly significant, but it shows the power of our method in the case of high-precision experimental data.

We verified that the influence of the specific form of the phase $\phi$ is, as expected, small.

The calculation presented here points out toward a real new effect revealed by our method.
This new effect might simply mean that $\rho$ is not a constant but a function of $t$, as well as $B$ might not be a constant but also a function of $t$.
In others words one must make the analysis of the experimental data with more sophisticated analytic forms of the scattering amplitude that the exponential one.

The restricted range (\ref{new22}) of values of $t_{min}$ obtained from our analysis
 is explicitly shown in Fig.~4, where we plot $\Delta_R^{exp}(\sqrt{s}=52.8\mbox{ GeV},t_i)$ computed from a model dynamically different from the exponential form, the Gauron-Leader-Nicolescu (GLN) model \cite{gau88}. 
This model builds the scattering amplitudes from the asymptotic theorems constraints as a combination of Bessel functions and Regge forms, embodies the Heisenberg-Froissart $\ln^2s$ behavior for $\sigma_T$ and includes the maximal Odderon \cite{luk73}.
In this case, $\rho(s,t)$ at a given $s$ is no more a constant but varies with $t$.
This dynamical characteristics are translated through the fact that $\Delta_R^{GLN}$, as it can be seen from Fig. 4, has a \textit{fast increase} in the region (\ref{new16}), in agreement with the increase shown by $\Delta_R^{exp}$.
This fast increase shows the importance of $ReF_N$ in the GLN model, in contrast with the exponential model.


The value of $t_{min}^{pp}$, extracted from $\Delta_R^{exp}$ by using $ImF_N$ as given by the GLN model, is perfectly compatible with the value (\ref{new22}).
A problem still persists: the value of $t_{min}^{pp}$, extracted from $\Delta_R^{th}$, is $-0.016 \mbox{ GeV}^{-2}$, in disagreement with the value (\ref{new22}).

The disagreement between $\Delta_R^{th}$ and $\Delta_R^{exp}$ is seen also through the values of 
$\chi^2/pt$.
The overall $\chi^2/pt$ value is comparable with the one in the exponential model case:
2.3/pt for a total of 34 points.
Again, the major contribution comes from the last 10 points.

It has to be noted that the GLN model has a much richer dynamical content than the exponential model, both from theoretical and phenomenological points of view. Moreover, it fits a large number of data for $pp$ and 
$\bar pp$ scattering in a huge range of $s$ ($4.5\le\sqrt{s}\le 541$ GeV) values, while the exponential parameters are fixed from fits performed scattering by scattering and energy by energy.

We conclude that neither the exponential model nor the GLN model can reproduce entirely the effect discussed in the present paper : the disagreement between $\Delta_R^{th}$ and $\Delta_R^{exp}$. 
However, the \textit{stability} of the value $t_{min}^{pp}$ extracted from $\Delta_R^{exp}$ is remarkable: in both models examined in the present paper this value is perfectly compatible with the value (\ref{new22}).

There are yet not $pp$ data at LHC.
However we can evaluate $\Delta_R^{th}$ from Eq.~(\ref{Del}) by assuming an exponential form (\ref{new15}) for $ReF_N$, e.g. with a slope $B=22\ (\mathrm{GeV})^{-2}$, $\sigma_T=111.5$~mb \cite{cud02} and with $\rho=0.15$ as illustrative values (see Fig. 5).
One gets a zero in $\Delta_R^{th}$ located at $t_{min}=-0.0044\ (\mathrm{GeV})^2$.
The future small-$t$ experiments at LHC \cite{efth} may detect the zero in  $\Delta_R^{exp}$ leading to the computation of $ReF_N^{pp}$ at $t_{min}^{pp}$ in terms of the Coulomb amplitude.
This would provide a strong constraint on the parameter $\rho(\sqrt{s}=14\mbox{ TeV, }t=0)$.
This constraint is crucial in detecting new phenomena in strong interactions (e.g., the Odderon presence).


\section{Conclusions}
In conclusion, we did find a new method for the determination of the real part of the elastic proton-proton amplitude at high $s$ and small $t$ at a given point $t_{min}^{pp}$ near $t=0$.
The real part of the hadron amplitude is computed, at $t=t_{min}^{pp}$, from the known Coulomb amplitude.

There are no hidden assumptions: we use data for $dN/dt$ and a given form of $ImF_N^{pp}$.
The usual method obviously needs to formulate, in addition, a given model for $ReF_N^{pp}$.

Our  method provides a powerful consistency check for the existing models and data and has a predictive power for the future measurements of the $\rho$-parameter at LHC.
It requires high-precision data and a large number of experimental points.
We illustrated how our method works by using the data at $\sqrt{s}=52.8$ GeV (Ref. \cite{528}).

As a byproduct of our method we discovered two new effects in the data at $\sqrt{s}=52.8$ GeV:
1. the significant discrepancy between $\Delta_R^{th}$ as defined in Eq. (\ref{Del}) and $\Delta_R^{exp}$ as defined in Eq. (\ref{Del2}), $\Delta_R^{th}$ involving $ReF_N$ while $\Delta_R^{exp}$ involves $ImF_N$; 2. $\Delta_R^{exp}$ goes through a minimum around a $t$-value $\vert t\vert\simeq 0.030-0.035\mbox{ GeV}^{-2}$ and has a sharp increase after this $t$-value (see Figs. 3-4).

The dynamical origin of these general effects is still obscure.
Maybe they are a result of oscillations in the very small $t$ region. 
In order to clarify their dynamical origin, high-precision experimental data at a high energy other than  $\sqrt{s}=52.8$ GeV are needed.
In principle, the experiments which will be performed at LHC \cite{efth} could explore this problem.

Stimulated by our findings, Kundrat and Lokajicek \cite{kl} tried recently
(six months after the publication of our results in a preprint form) to
generalize our method at higher t-values. These authors write that the
existence of a rather sharp minimum in our approach ``has provoked'' them
``to perform a more detailed analysis in this region with the help of
general eikonal approach''. Unfortunately, they add that our results are
``burdenened by two decisive discrepancies: non-allowed renormalization of
experimental data and application of internally inconsistent simplified
approach of West and Yennie''. This assertion is unfair, because: 1. our
results are independent of any renormalization of the data; 2. the extension
of the standard Coulomb-nuclear phase for all the range of t-values is
beyond the scope of the present paper. Moreover, as one can see from Fig. 2
of Ref. \cite{kl} and from the comments of the authors on this figure, the
sharp minimum in their generalized $\Delta_R$ is get precisely when our
equation (12) is satisfied and its locations, for peripheral and central
behaviours, exactly correspond to our numerical results for the two models
which we studied. In fact, the supplementary term proportional with $\alpha$, induced in our equation (10) by the eikonal model of Ref. \cite{kl} (see
their equation (29)), produces negligible changes in the region of very
small t and, therefore, our results are not significantly affected by the
generalized formalism of Ref. \cite{kl}.

Let us note that our method can be easily extended (with minor changes) to proton-antiproton scattering, by observing that, in this case, it is the combination
\begin{equation}
\label{comb}
ReF_N^{\bar pp}- F_C^{\bar pp}
\end{equation}
which must go through zero at some value $t=t_{min}^{\bar pp}$.
The method described in the present paper could be therefore used to analyze the UA4 data at $\sqrt{s}=541$ GeV \cite{ua42}, a complex work which will be done and presented in a separate paper.
Of course, in general, one expects that $t_{min}^{pp}\neq t_{min}^{\bar pp}$ at fixed $s$.

Our method could be also extended to the case of proton-nucleus scattering at high energies.

\textit{Acknowledgments.} The authors thank Dr. C. Bourrely and Dr. J. Soffer for useful discussions. One of us (O.V.S.) thanks Dr. Jean-Eudes Augustin for hospitality at LPNHE Paris, where most of the present work was done.

\newpage
\noindent \textbf{Figure captions}
\vspace{1cm}

\noindent\textbf{Fig.~1}
$\Delta_R^{th}$ (the solid curve) and $\Delta_R^{exp}$ (the triangle points)  for  $pp$ scattering 
(Eqs. (\ref{Del}) and (\ref{Del2})) at $\sqrt{s}=52.8$ GeV  as a function of $t$, computed with the exponential form of the amplitude (Eqs. (\ref{im}) and (\ref{new15})).

\vspace{1cm}

\noindent\textbf{Fig.~2}
The ratio $r$ (see Eq. (\ref{new19})) for $pp$ scattering at $\sqrt{s}=52.8$ GeV  as a function of $t$, where $d\sigma^{th}/dt$ is computed by using the exponential form of the amplitude (Eqs. (\ref{im}) and (\ref{new15}).

\vspace{1cm}

\noindent\textbf{Fig.~3}
$\Delta_R^{exp}$ (the triangle points)  for  $pp$ scattering (Eq. (\ref{Del2})) at $\sqrt{s}=52.8$ GeV  as a function of $t$, computed with the exponential form of $ImF_N$ (Eq. (\ref{im})) and fitted by the polynomial form (\ref{new20}) - (\ref{new21}) (the solid curve).
The arrow indicates the position of $t_{min}^{pp}$.

\vspace{1cm}

\noindent\textbf{Fig.~4}
$\Delta_R^{th}$ (the solid curve) and $\Delta_R^{exp}$ (the triangle points)  for  $pp$ scattering 
(Eqs. (\ref{Del}) and (\ref{Del2})) at $\sqrt{s}=52.8$ GeV  as a function of $t$, computed within the Gauron-Leader-Nicolescu (GLN) model (Ref. \cite{gau88}).

\vspace{1cm}

\noindent\textbf{Fig.~5}
$\Delta_R^{th}$  for  $pp$ scattering (Eq. \ref{Del}) at $\sqrt{s}=14$ TeV  computed within the exponential model (Eqs. (\ref{im}) and (\ref{new15})) with the illustrative values $B=22$ (GeV)$^{-2}$, $\sigma_T=111.5$ mb and $\rho=0.15$.

\newpage

\begin{center}
\includegraphics{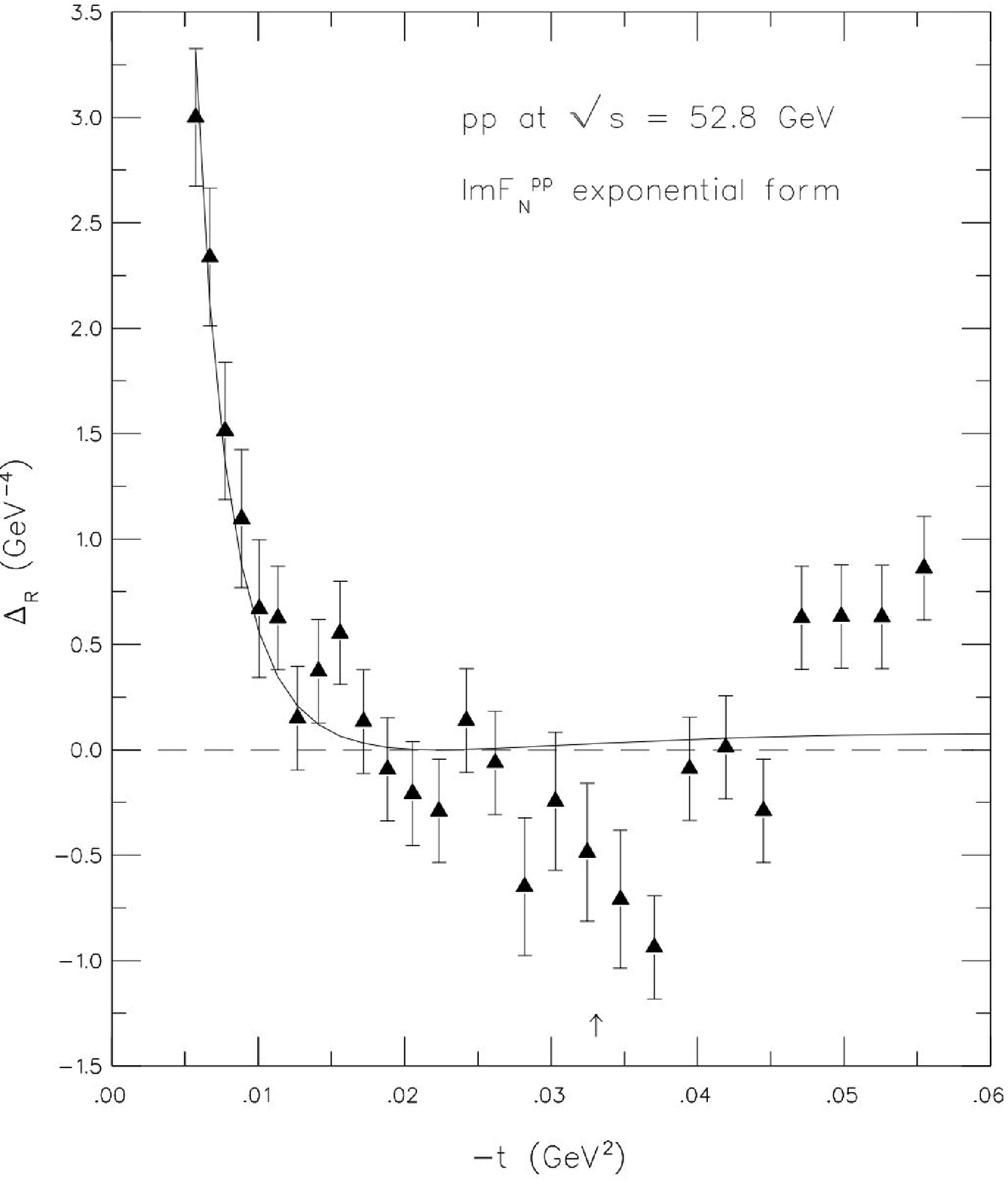}
{\LARGE Fig. 1}
\end{center}

\newpage
\begin{center}
\includegraphics{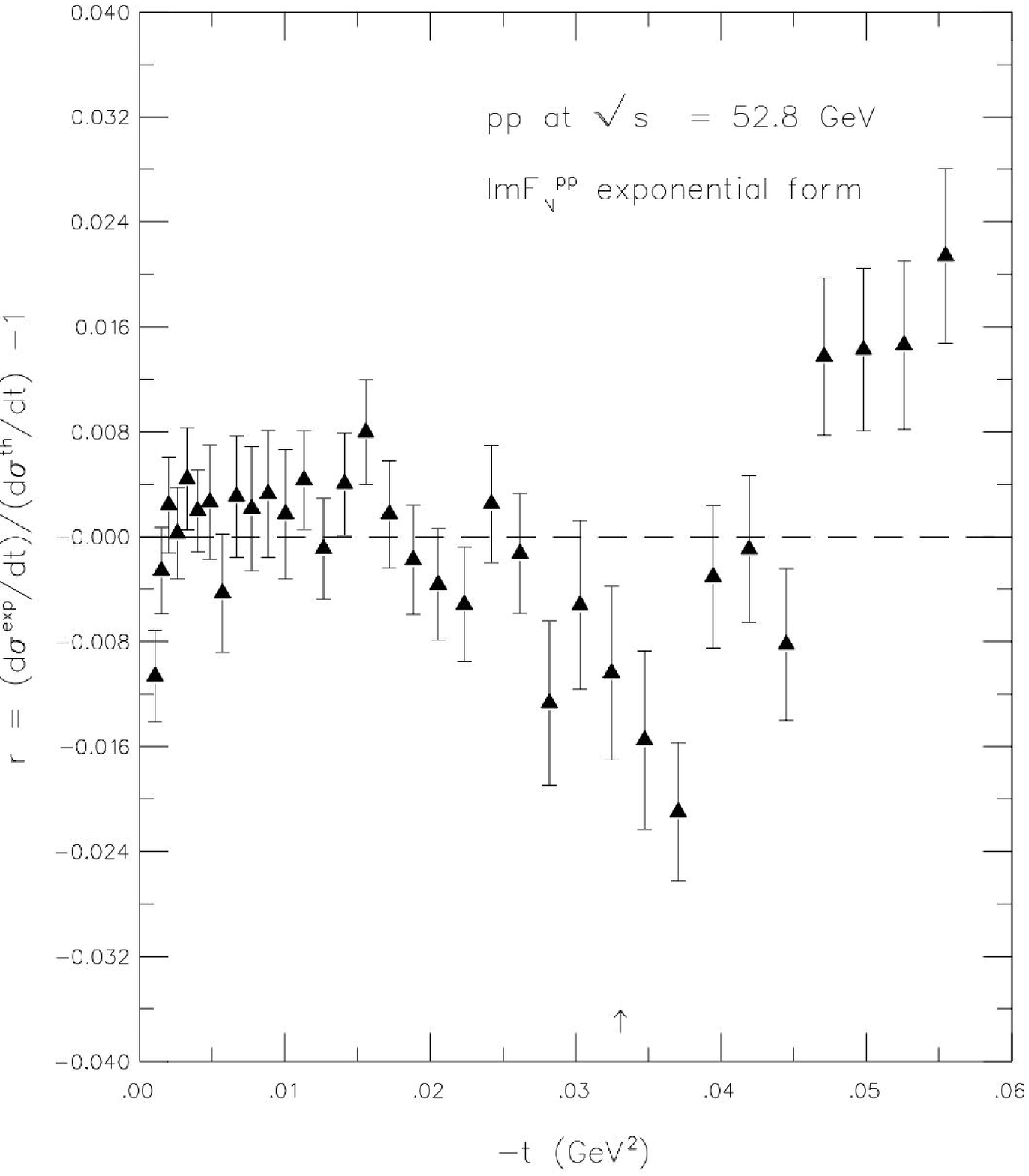}
{\LARGE Fig. 2}
\end{center}
\newpage
\newpage
\begin{center}
\includegraphics{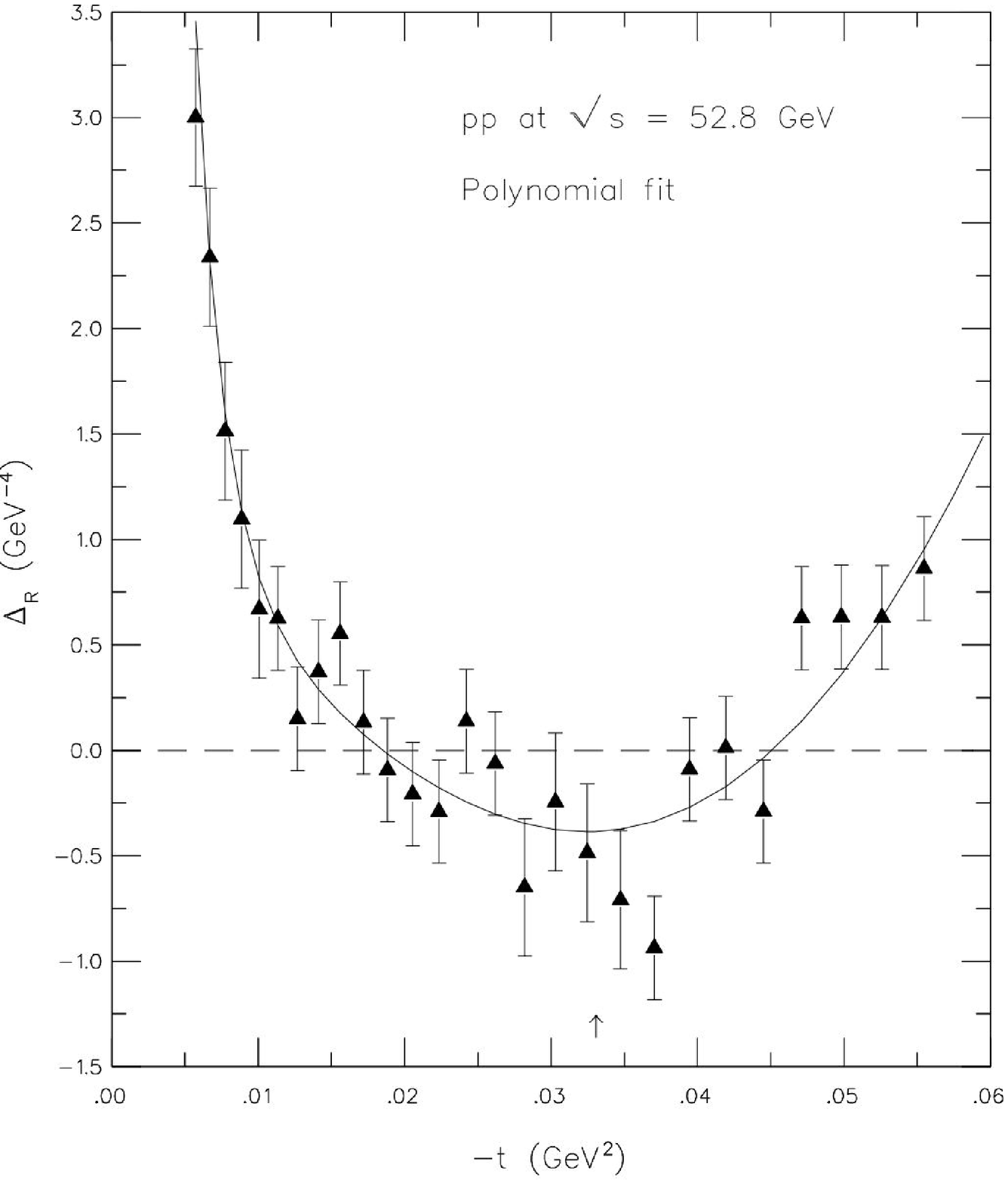}
{\LARGE Fig. 3}
\end{center}

\newpage
\begin{center}
\includegraphics{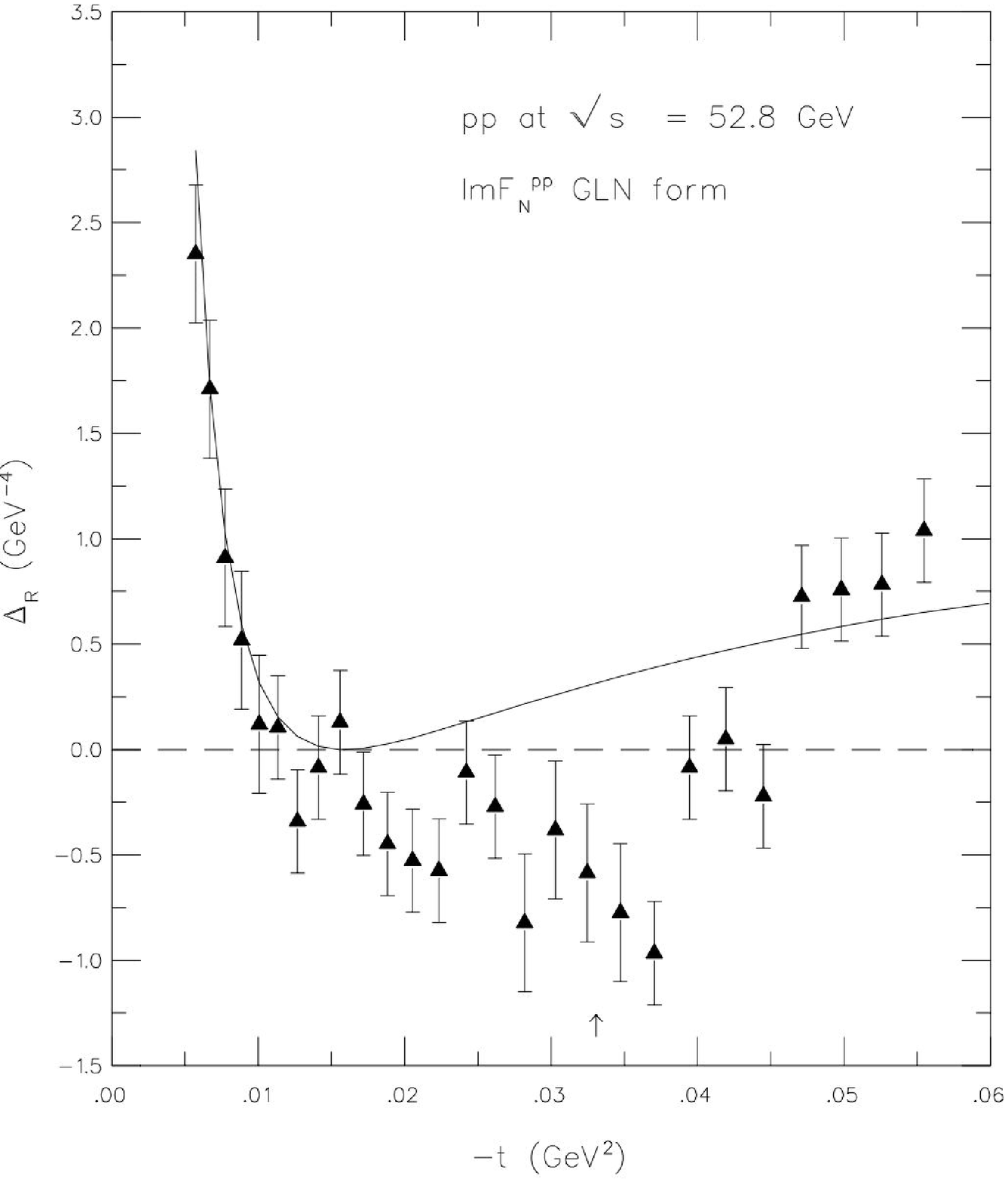}
{\LARGE Fig. 4}
\end{center}
\newpage

\newpage
\begin{center}
\includegraphics{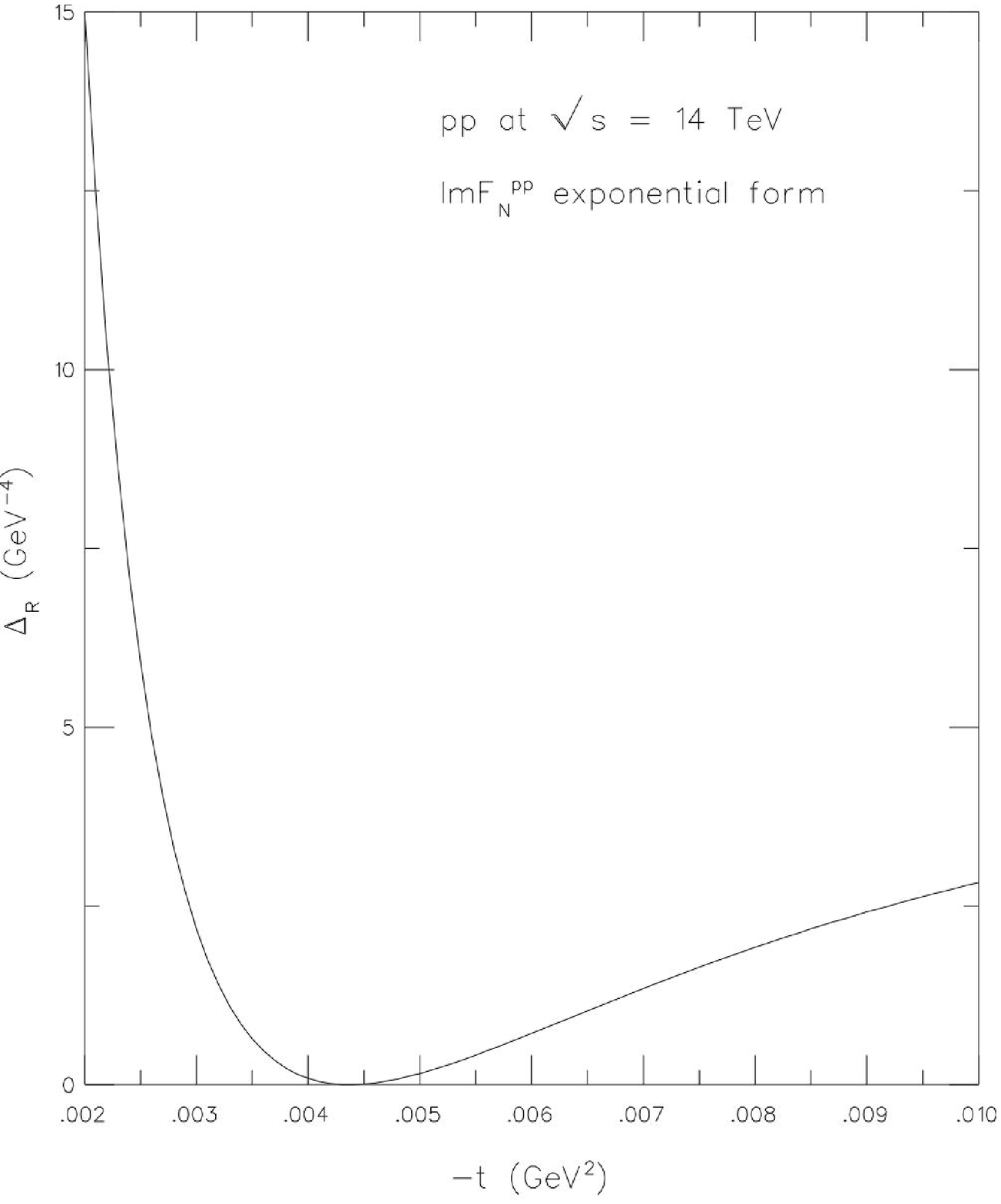}
{\LARGE Fig. 5}
\end{center}
\newpage



 \end{document}